\documentclass[12pt]{article}
\usepackage{graphicx}
\usepackage{epsfig}
\usepackage{appendix}
\begin{document}
\title{ Isgur-Wise function in a QCD inspired potential model with confinement as parent in the  Variationally Improved Perturbation Theory (VIPT) }
\author{$^{1}$Bhaskar Jyoti Hazarika and $^{2}$D K Choudhury \\
$^{1}$Dept of Physics,Pandu College,Guwahati-781012,India\\
e-mail:bh53033@gmail.com\\
$^{2}$Dept. of Physics, Gauhati University, Guwahati-781014,India}
\date{}
\maketitle
\begin{abstract}
We have recently reported the calculation of slope and curvature of Isgur-Wise function based on Variationally Improved Perturbation Theory (VIPT) in a QCD inspired potential model. In that work, Coulombic potential was taken as the parent while the linear one as the perturbation.In this work, we choose the linear one as the parent with Coulombic one as the perturbation and see the consequences.  
\end{abstract}
\section{Introduction}
The use of VIPT \cite{1} in the calculation of slope and curvature of  Isgur-Wise(I-W) function is very recent.In a QCD inspired potential model , we have two options in using VIPT \cite{2,3,4} with the linear cum Coulomb potential of QCD\cite{18} -(i) Coulombic one as the parent, the linear one as the perturbation and (ii) linear one as the parent,Coulombic one as the perturbation.Already,Coulombic part as parent has been analyzed\cite{1} succesfully taking into account of three terms in the summation of equation(10) of Ref 1 for  $D, D_{s} and  B$  mesons .However, the results were shown to be improved only with more terms in the equation(10) of Ref 1.In this work, we take an alternate strategy : we take the linear potential as the parent and see any modification over the results of earlier work\cite{1} for the same three number of terms of equation(10)in that work. We recall that \cite{3} for the linear potential to be dominant we prefer  $<r>$    $ >$    $r_{0}$ ,where $<r>$ is the expectation value of the distance $r$ which reasonably gives the size of a state(in this case meson) and $r_{0}$ is a point at which linear cum Coulomb potential becomes zero( Fig.1 of Aitchison and Dudek ,Ref [3] ).In this work we have also included the $B_{s} and  B_{c}$ mesons which have greater reduced mass $\mu$, to check the range of  applicability of the approach.\\

The rest of the paper is organized as follows : section 2 contains the formalism, section 3 the result and calculation while section 4 includes the discussion and conclusion. \\
   
\section{Formalism}
\subsection{Isgur-Wise function; its slope and curvature}
The Isgur-Wise function is written as \cite{5} :
\begin{eqnarray}
\xi\left(v_{\mu}.v^{\prime}_{\mu}\right)\nonumber&=&\xi\left(y\right)\\&=&1-\rho^{2}\left(y-1\right)+ C\left(y-1\right)^{2}+...
\end{eqnarray}
where 
\begin{equation}
y= v_{\mu}.v^{\prime}_{\mu}
\end{equation}
and $v_{\mu}$ and $v^{\prime}_{\mu}$  being the four velocity of the heavy meson before and after the decay.The quantity $\rho^{2}$  is the slope of I-W function at $y=1$ and known as charge radius :
\begin{equation}
\rho^{2}= \left. \frac{\partial \xi}{\partial y}\right.|_{y=1}
\end{equation}
The second order derivative is the curvature of the I-W function known as convexity parameter :
\begin{equation}
C=\left .\frac{1}{2}\right. \left(\frac{\partial^2 \xi}{\partial y^{2}}\right)|_{y=1}
\end{equation}
For the heavy-light flavor mesons the I-W function can also be written as \cite{8,17} :
\begin{equation}
\xi\left(y\right)=\int_{0}^{+\infty} 4\pi r^{2}\left|\psi\left(r\right)\right|^{2}\cos pr dr
\end{equation}
where
\begin{equation}
p^{2}=2\mu\left(y-1\right)
\end{equation}
Now the wavefunction $\psi$  of the hadronic system is determined by taking the linear as parent potential.\\

\subsection{First order corrected wavefunction and energy in VIPT}

 The wavefunction corrected upto the first order of $j^{th}$ state  is given by (Equation 10 of Ref 1) :                                                                             \begin{equation}                                                             \psi_{j}= \psi_{j}^{(0)}+\sum_{k\neq j}\frac{\int\psi_{k}^{(0)*}H_{P^{\prime}j}^{\prime}\psi_{j}^{(0)}dv}{E_{j}^{(0)}-E_{k}^{(0)}}
                      \end{equation}                                                              The energy corrected upto first order for the same state is :                     \begin{eqnarray}                                                              E_{j}\nonumber&=&\int \psi_{j}^{(0)*} H\psi_{j}^{(0)} dv\\&=&\int \psi_{j}^{(0)*}( H_{oP^{\prime}}+H_{P^{\prime}}^{\prime}) \psi_{j}^{(0)} dv 
               \end{eqnarray}                                                                where $\psi_k$, $E_k$ are the wavefunction and energy eigen values of the $k^{th}$ states which are orthonormal to $j^{th}$ state .The superscript$(0)$ means zeroeth order correction of the corresponding quantities .Also , we note that  $P^{\prime}$is the variational parameter and  $H_{oP^{\prime}}$ , $ H_{P^{\prime}}^{\prime}$ are as defined in equation(9)of Ref 1. \\ 
 
                                                                                               The summation in equation (7) above [Equation 10 of Ref 1] , can include any number of $ k^{th}$ states .In this work , we consider terms upto three states in the summation as was done Ref[1].\\
    
 \subsection{ Wavefunctions using VIPT with linear potential as the parent}            \subsection*{(i)With one term in the summation}                                     As explained earlier , we take $b^{\prime}$  as the variational parameter  instead of the physical parameter $b$ in the parent linear potential to write the Hamiltonian as\cite{10,7}:                       \begin{eqnarray}                                                                H\nonumber&=&H_{o}+H^{\prime}\\\nonumber&=&-\frac{\nabla^{2}}{2\mu}+br-\frac{4\alpha_s}{3r}+c\\\nonumber&=&-\frac{\nabla^{2}}{2\mu}+br-\frac{\alpha}{r}+c\\\nonumber&=&-\frac{\nabla^{2}}{2\mu}+b^{\prime}r-\frac{\alpha}{r}-b^{\prime}r +br+c\\&=&H_{o b^{\prime}}+H_{b^{\prime}}^{\prime}  
                     \end{eqnarray}                                                                where                                                                          $\alpha=\frac{4\alpha_s}{3}$                              Now ,$H_{o b^{\prime}}=-\frac{\nabla^{2}}{2\mu}-b^{\prime}r$ is the parent Hamiltonian with the new parameter $b^{\prime}$ and $H_{b^{\prime}}^{\prime}=\frac{\alpha}{r}-b^{\prime}r+br+c$ is the perturbed Hamiltonian with the same variational parameter $b^{\prime}$ instead of the physical parameter $b$ .\\ 
                                                     
 We consider $j^{th}$ as $1s$ state ($n=1$,$l=0$) and in the summation of equation(7)[ equation(10) of Ref 1], we consider a single  $k^{th}$state which is the $2s$ state ($n=2$,$l=0$).\\
 
We note that in the variational method, we are interested only in the $'r'$ dependence of the Hamiltonian, and so $'c'$ in $H_{b^{\prime}}^{\prime}$ has no role to play in the calculation\cite{9}.\\

The unperturbed wavefunctions with linear parent with appropriate boundary conditions are the Airy functions given by \cite{3}:
\begin{equation}
\psi_{n0}\left(r\right)=\frac{N_{n}}{2\sqrt{\pi} r}Ai\left((2\mu b^{\prime})^{\frac{1}{3}}r +\rho_{on}\right)
\end{equation}
 where $\rho_{0n}$ s are the zeroes of the Airy function $Ai\left(\rho_{on}\right)=0$ given by \cite{6}:
\begin{equation}
\rho_{on}=-\left[\frac{3\pi(4n-1)}{8}\right]^{\frac{2}{3}}
\end{equation}
and $N_{n}$ is the normalization constant.\\

As an illustration , we reproduce for $s$ states a few of the zeroes of the Airy function in table 1.\\

\begin{table}
\begin{center}
\caption{A few of the zeroes of Airy function for $s$ states .}
\begin{tabular}{|c|c|}\hline
State&$\rho_{0n}$\\\hline
$1s$&-2.3194\\\hline
$2s$&-4.083\\\hline
$3s$&-5.5183\\\hline
$4s$&-6.782\\\hline
\end{tabular}
\end{center}
\end{table}
 The corresponding energies are  given as :
\begin{equation}
E_{n}=-\left(\frac{b^{\prime^{2}}}{2\mu}\right)^{\frac{1}{3}}\rho_{0n}
\end{equation}
Of course $n=1,2,3,4,.....$ is the principal quantum number.\\

          Thus  the trial $1s$ state ($n=1,l=0$) wavefunction is (which is also the unperturbed wavefunction) :                                                                                  \begin{eqnarray}                                                              \psi^{(0)}\nonumber&=& \psi_{10}^{(0)}\\\nonumber&=&\frac{N_{1}}{2\sqrt{\pi} r}Ai\left((2\mu \overline{b}^{\prime})^{\frac{1}{3}}r -2.3194\right)\\&=&\frac{N_{1}}{2\sqrt{\pi} r}Ai\left(z_{1}\right)
         \end{eqnarray}
where 
\begin{equation}
z_{1}= \left((2\mu \overline{b}^{\prime})^{\frac{1}{3}}r -2.3194\right)
\end{equation}
and the subscript $10$   indicates the quantum number ($n,l$)  of the $j^{th}$  state.\\

We note that $b^{\prime}$ is replaced by $\overline{b}^{\prime}$ which is obtained by minimizing $E_{j}$ given by equation(8).It is essential since in VIPT  we have to use the values of variational parameter leading to minimum energy (for example in Ref 1, $\alpha_{s}$ was replaced by $\overline{\alpha}_{10}^{\prime}$). The values of $\overline{b}^{\prime}$ for different mesons are listed in table 2.\\

Now we consider the single $ k^{th}$ state in the summation of equation (7)which is the $2s$ state given by :
 \begin{eqnarray}                                                              \ \psi_{20}^{(0)}\nonumber&=&\frac{N_{2}}{2\sqrt{\pi} r}Ai\left((2\mu \overline{b}^{\prime})^{\frac{1}{3}}r -4.083\right)\\&=&\frac{N_{2}}{2\sqrt{\pi} r}Ai\left(z_{2}\right)
         \end{eqnarray}
where 
\begin{equation}
z_{2}= \left((2\mu \overline{b}^{\prime})^{\frac{1}{3}}r -4.083\right)
\end{equation}  
The wavefunction corrected upto first order is :
\begin{equation}
\psi_{S}=N\left[\psi^{(0)}+\frac{\left(2\mu\right)^{\frac{1}{3}}}{\left(\rho_{02}-\rho_{01}\right)\overline{b}^{\prime^{\frac{2}{3}}}}\left(\left(b-\overline{b}^{\prime}\right)<r>_{2,1}-\alpha<\frac{1}{r}>_{2,1}\right)\psi_{20}\left(r\right)\right]
\end{equation}
where
\begin{equation}
<r>_{2,1}=N_{1}N_{2}\int_{0}^{+\infty}Ai\left((2\mu \overline{b}^{\prime})^{\frac{1}{3}}r -2.3194\right)Ai\left((2\mu \overline{b}^{\prime})^{\frac{1}{3}}r -4.083\right)
\end{equation}
and $N$ is the normalization constant.\\

\subsection*{(ii)With two terms in the summation}
 We next consider the $3s$ state ($n=3$,$l=0$) in addition to $2s$ state (as done in the single term case)given by :
 \begin{eqnarray}                                                              \ \psi_{30}^{(0)}\nonumber&=&\frac{N_{3}}{2\sqrt{\pi} r}Ai\left((2\mu \overline{b}^{\prime})^{\frac{1}{3}}r -5.5153\right)\\&=&\frac{N_{3}}{2\sqrt{\pi} r}Ai\left(z_{3}\right)
         \end{eqnarray}
where 
\begin{equation}
z_{3}= \left((2\mu \overline{b}^{\prime})^{\frac{1}{3}}r -5.5153\right)
\end{equation} 

With the inclusion of this state , the wavefunction corrected upto the first order is :
\begin{eqnarray}
\psi_{D}=N^{\prime}[\psi^{(0)}+\nonumber\frac{\left(2\mu\right)^{\frac{1}{3}}}{\left(\rho_{02}-\rho_{01}\right)\overline{b}^{\prime^{\frac{2}{3}}}}\left(\left(b-\overline{b}^{\prime}\right)<r>_{2,1}-\alpha<\frac{1}{r}>_{2,1}\right)\psi_{20}\left(r\right)+
\\ \frac{\left(2\mu\right)^{\frac{1}{3}}}{\left(\rho_{03}-\rho_{01}\right)\overline{b}^{\prime^{\frac{2}{3}}}}\left(\left(b-\overline{b}^{\prime}\right)<r>_{3,1}-\alpha<\frac{1}{r}>_{3,1}\right)\psi_{30}\left(r\right)]
\end{eqnarray}
where
\begin{equation}
<r>_{3,1}=N_{1}N_{3}\int_{0}^{+\infty}Ai\left((2\mu \overline{b}^{\prime})^{\frac{1}{3}}r -2.3194\right)Ai\left((2\mu \overline{b}^{\prime})^{\frac{1}{3}}r -5.5153\right)
\end{equation}
and $N^{\prime}$ is the normalization constant.\\

\subsection*{(iii)With three terms in the summation}
In addition to the $2s$ and $3s$ states we now add the $4s$ state : 

 \begin{eqnarray}                                                              \ \psi_{40}^{(0)}\nonumber&=&\frac{N_{4}}{2\sqrt{\pi} r}Ai\left((2\mu \overline{b}^{\prime})^{\frac{1}{3}}r -6.782\right)\\&=&\frac{N_{3}}{2\sqrt{\pi} r}Ai\left(z_{4}\right)
         \end{eqnarray}
where 
\begin{equation}
z_{4}= \left((2\mu \overline{b}^{\prime})^{\frac{1}{3}}r -6.782\right)
\end{equation} 
With the inclusion of this state , the first order wavefunction now becomes :

\begin{eqnarray}
\psi_{T}=N^{\prime\prime}[\psi^{(0)}+\nonumber\frac{\left(2\mu\right)^{\frac{1}{3}}}{\left(\rho_{02}-\rho_{01}\right)\overline{b}^{\prime^{\frac{2}{3}}}}\left(\left(b-\overline{b}^{\prime}\right)<r>_{2,1}-\alpha<\frac{1}{r}>_{2,1}\right)\psi_{20}\left(r\right)+
\\\nonumber \frac{\left(2\mu\right)^{\frac{1}{3}}}{\left(\rho_{03}-\rho_{01}\right)\overline{b}^{\prime^{\frac{2}{3}}}}\left(\left(b-\overline{b}^{\prime}\right)<r>_{3,1}-\alpha<\frac{1}{r}>_{3,1}\right)\psi_{30}\left(r\right)\\
+\frac{\left(2\mu\right)^{\frac{1}{3}}}{\left(\rho_{04}-\rho_{01}\right)\overline{b}^{\prime^{\frac{2}{3}}}}\left(\left(b-\overline{b}^{\prime}\right)<r>_{4,1}-\alpha<\frac{1}{r}>_{4,1}\right)\psi_{40}\left(r\right)]
\end{eqnarray}
where
\begin{equation}
<r>_{4,1}=N_{1}N_{4}\int_{0}^{+\infty}Ai\left((2\mu \overline{b}^{\prime})^{\frac{1}{3}}r -2.3194\right)Ai\left((2\mu \overline{b}^{\prime})^{\frac{1}{3}}r -6.782\right)
\end{equation}
and $N^{\prime\prime}$ is the normalization constant.\\

 The relativistic version of these wavefunctions are obtained by multiplying above expression by $(r\mu\alpha)^{-\epsilon}$ \cite{11,12}.  Thus,relativistic version of all these wavefunctions is:
\begin{equation}
\psi_{i,rel}=\psi_{i}\left(r\mu\alpha\right)^{-\epsilon}
\end{equation}

where $i=S,D,T$\\

and 
\begin{equation}
\epsilon =1-\sqrt{1-\frac{4\alpha_{s}}{3}}
\end{equation}

Putting all these wavefunctions i.e. equations (17),(21),(25)  and (27) in (5) we can calculate the Isgur-Wise function for the different cases.\\

\section{Calculation and Results}

We have listed  the  values of charge radius and convexity parameter of the calculated I-W function for various heavy-light flavor mesons in the present method considering  single state , two states,and three states in the summation occurred in VIPT with perturbative  Coulombic  and  relativistic effect.\\

In tables(3-5), we record the predictions of the present work for single term,two terms and three terms respectively .In table 6, we give comparision of this work to that of Ref[1]; while table 7 gives a summary of the predictions of slope and curvature of IW function in other models.\\

  The  $\alpha_{s}$ values are taken from the  $V$-scheme \cite{13,14,15,16} and the integrations are done numerically for all these calculations.\\

\begin{table}
\begin{center}
\caption{Values of $\overline{b}^{\prime}$ with $b=0.183GeV^{2}$.}
\begin{tabular}{|c|c|c|c|c|}\hline
Mesons&Reduced mass $\mu$&$\alpha=\frac{4\alpha_{s}}{3}$&$\overline{b}^{\prime}$without
 relativistic effect&$\overline{b}^{\prime}$with
 relativistic effect\\\hline
$D$&0.2761&0.924&5.406&16.24\\\hline
$D_{s}$&0.368248&0.924&5.876&19.8\\\hline
$B$&0.31464&0.348&4.33&5.587\\\hline
$B_{s}$&0.4401&0.348&4.497&5.954\\\hline
$B_{c}$&1.1803&0.348&5.39&8.103\\\hline

\end{tabular}
\end{center}
\end{table}

\begin{table}
\begin{center}
\caption{Values of slope $\rho^{2}$ and curvature $C$ with single term in equation(7) }
\begin{tabular}{|c|c|c|c|c|}\hline
Meson&$\rho_{S}^{2}$&$C_{S}$&$\rho_{S,rel}^{2}$&$C_{S,rel}$\\\hline
$D$&1.36&0.01&0.53&0.0022\\\hline
$D_{s}$&1.867&0.03&0.702&0.0036\\\hline
$B$&1.93&0.02&1.41&0.013\\\hline
$B_{s}$&2.923&0.046&2.113&0.0283\\\hline
$B_{c}$&9.442&0.484&6.274&0.2522\\\hline

\end{tabular}
\end{center}
\end{table}

\begin{table}
\begin{center}
\caption{Values of slope $\rho^{2}$ and curvature $C$ with two terms in equation(7) }
\begin{tabular}{|c|c|c|c|c|}\hline
Meson&$\rho_{D}^{2}$&$C_{D}$&$\rho_{D,rel}^{2}$&$C_{D,rel}$\\\hline
$D$&1.201&0.013&0.57&0.0026\\\hline
$D_{s}$&2.001&0.0242&0.74&0.0041\\\hline
$B$&2.004&0.0244&1.44&0.0133\\\hline
$B_{s}$&3.031&0.0565&2.16&0.0297\\\hline
$B_{c}$&10.2&0.61&6.51&0.275\\\hline

\end{tabular}
\end{center}
\end{table}
\begin{table}
\begin{center}
\caption{Values of slope $\rho^{2}$ and curvature $C$ with three terms in equation(7) }
\begin{tabular}{|c|c|c|c|c|}\hline
Meson&$\rho_{T}^{2}$&$C_{T}$&$\rho_{T,rel}^{2}$&$C_{T,rel}$\\\hline
$D$&1.33&0.016&0.604&0.00326\\\hline
$D_{s}$&2.023&0.0305&0.78&0.0054\\\hline
$B$&2.027&0.031&1.54&0.0217\\\hline
$B_{s}$&3.087&0.071&2.29&0.047\\\hline
$B_{c}$&10.25&0.767&6.99&0.441\\\hline

\end{tabular}
\end{center}
\end{table}
\begin{table}
\caption{ Comparision of this work to that of Ref[1] with  relativistic effect being included. }
\begin{tabular}{|r|r|r|r|r|r|r||r|r|r|r|r|r|}\hline
Me-&\multicolumn{6}{c||}{Ref[1]}&\multicolumn{6}{c|}{This work}\\
\cline{2-13}
son  &\multicolumn{1}{r|}{$\rho_{S}^{2}$}&$C_{S}$&$\rho_{D}^{2}$&$C_{D}$&$\rho_{T}^{2}$&$C_{T}$ &\multicolumn{1}{r|}{$\rho_{S}^{2}$}&$C_{S}$&$\rho_{D}^{2}$&$C_{D}$&$\rho_{T}^{2}$&$C_{T}$\\\hline 

$D$&\multicolumn{1}{r|}{.433}&.525&.432&.524&.43&.516&\multicolumn{1}{r|}{0.53}&0.0022&0.57&0.0026&0.604&0.0033\\
& & & & & & & & & & & &\\\hline
$B$&\multicolumn{1}{r|}{.56}&.85&.55&.84&.545&.815&\multicolumn{1}{r|}{0.702}&0.0036&0.74&0.0041&0.78&0.0054\\
& & & & & & & & & & &&\\\hline
$D_{s}$&\multicolumn{1}{r|}{3.6}&15.3&3.16&12.32&3.12&11.8&\multicolumn{1}{r|}{1.41}&0.0126&1.44&0.0133&1.54&0.0213\\
& & & & & & & & & &&&\\\hline
$B_{s}$&\multicolumn{1}{r|}{...}&...&...&...&...&...&\multicolumn{1}{r|}{2.11}&0.0283&2.16&0.0297&2.29&0.0471\\
& & & & & & & & & &&&\\\hline
$B_{c}$&\multicolumn{1}{r|}{...}&...&...&...&...&...&\multicolumn{1}{r|}{6.27}&0.252&6.51&0.275&6.99&0.441\\
& & & & & & & & & &&&\\\hline
\end{tabular}
\end{table}

\begin{table}
\begin{center}
\caption{Predictions of the slope and curvature of the IW function in various models.}
\begin{tabular}{|c|c|c|}\hline
Model& Value of $\rho^{2}$ &Value of curvature $C$\\\hline
Le Youanc et al \cite{19}&$\ge 0.75$&..\\
Le Youanc et al \cite{20}&$\ge 0.75$&$\ge 0.47$\\ 
Rosner \cite{28}&1.66&2.76\\
Mannel \cite{29,30}&0.98&0.98\\
Pole Ansatz \cite{31}&1.42&2.71\\
MIT Bag Model \cite{27}&2.35&3.95\\
Simple Quark Model \cite{26}&1&1.11\\
Skryme Model \cite{24}&1.3&0.85\\
QCD Sum Rule \cite{25}&0.65&0.47\\
Relativistic Three Quark Model \cite{23}&1.35&1.75\\
Infinite Momentum Frame Quark Model \cite{22}&3.04&6.81\\
Neubert \cite{32}&0.82$\pm$0.09&..\\\hline
\end{tabular}
\end{center}
\end{table}

\section{Discussion and Conclusion}
We note that in the present analysis , the slope and curvature increase with the inclusion of more terms in the summation of equation(7)in contrary to the case of Coulombic parent\cite{1}, where with more terms the slope and curvature decreased.An analysis of table 6 also indicate that for a definite term , the slope has increased in the present work compared to Ref[1],while for the curvature ,the pattern is reversed i.e. it has decreased in the present case compared to Ref[1].So,with only one term for the linear parent in equation(7),we can have comparable values for slope and curvature with other models and data [table-7].This is undoubtably a great phenomenological advantage as involvement of more terms in equation(7) makes the calculation quite cumbersome  which happened in Ref[1].However, relativistic correction in this case also decrease the slope and curvature of Isgur-Wise function as observed earlier \cite{1}.\\

To conclude , the present approach with the linear potential as the parent has provided much more improved results for the slope and specially curvature of the IW function within the QCD inspired potential model under study and appeared to be preferable over the one of Ref[1] where the linear potential was considered as perturbation.\\

\end{document}